\input{psfig}
\documentclass[12pt,preprint]{aastex}

\newcommand{\be}{\begin{equation}}
\newcommand{\ee}{\end{equation}}
\newcommand{\bea}{\begin{eqnarray}}
\newcommand{\eea}{\end{eqnarray}}
\shorttitle{ON THE AVERAGE COMOVING NUMBER DENSITY OF HALOS} \shortauthors{A. Del
Popolo}

\begin{document}


\title{The moving barrier and an
improved model for the multiplicity function
}


\author{A. Del Popolo\altaffilmark{1}}

\affil{Bo$\breve{g}azi$\c{c}i University, Physics Department,
     80815 Bebek, Istanbul, Turkey}

%\altaffiltext{2}{present address: Feza G\"ursey Institute, P.O. Box 6 \c Cengelk\"oy, Istanbul,
%     Turkey.}
\altaffiltext{2}{
Dipartimento di Matematica, Universit\`{a} Statale di Bergamo,
  Piazza Rosate, 2 - I 24129 Bergamo, ITALY
}


%\pagerange{\pageref{firstpage}--\pageref{lastpage}} \pubyear{2000}

\begin{abstract}

I compare the numerical multiplicity function given in Yahagi, Nagashima \& Yoshii (2004) with the theoretical multiplicity function obtained by means of the excursion set model and an improved version of the barrier
shape obtained in Del Popolo \& Gambera (1998), which implicitly takes account of
total angular momentum acquired by the proto-structure during evolution and of a non-zero cosmological constant.
I show that the multiplicity function obtained in the present paper, is in better agreement with Yahagi, Nagashima \& Yoshii (2004) simulations than other previous models (Sheth \& Tormen 1999; Sheth, Mo \& Tormen 2001; Sheth \& Tormen 2002; Jenkins et al. 2001) and that differently from some previous multiplicity function models (Jenkins et al. 2001; Yahagi, Nagashima \& Yoshii 2004)
 %The multiplicity function of the present paper gives a good fit to simulations results as the fit function proposed by %Yahagi, Nagashima \& Yoshii (2004), but differently from that
it was obtained from a
sound theoretical background.

\end{abstract}

%\begin{keywords}
%\end{keywords}

%\newpage

\section{Introduction}

Two different kinds of methods are widely used for the study of
the structure formation. The first one  is N-body simulations,
that are able to follow the evolution of a large number of
particles under the influence of the mutual gravity, from initial
conditions to the present epoch. The second one are
semi-analytical methods. Among them, Press-Schechter (hereafter PS) approach
and its extensions (EPS) are of great interest since they allow us
to compute mass functions (Press \& Schechter  1974; Bond et al.
1991), to approximate merging histories (Lacey \& Cole 1993, LC93
hereafter, Bower 1991, Sheth \& Lemson 1999b) and to estimate the
spatial clustering of dark matter haloes (Mo \& White 1996;
Catelan et al. 1998, Sheth \& Lemson 1999a).\\

Although the analytical framework of the PS model has been greatly refined and extended (Bond et al. 1991; Lacey and Cole 1993), it is well known that the PS mass function, while qualitatively correct, disagrees with the results of
N-body simulations. In particular, the PS formula overestimates the abundance of haloes near the characteristic mass
$M_{\ast}$ and underestimates the abundance in the high mass tail (Efstathiou et al. 1988;
%White, Efstathiou \& Frenk 1993;
Lacey \& Cole 1994; Tozzi \& Governato 1998; Gross et al. 1998; Governato et al. 1999).

A better agreement between the numerical mass function and the analytic mass function
%The quoted discrepancy can be
can be obtained by incorporating into the PS ansatz the non-sphericity of collapse model (Del Popolo \& Gambera 1998;
Sheth \& Tormen 1999 (hereafter ST) ; Sheth, Mo \& Tormen 2001 (hereafter SMT); Sheth \& Tormen 2002 (hereafter ST1); Jenkins et al. 2001 (hereafter J01)), instead of the spherical model or taking into account the spatial correlation of density fluctuations (Nagashima 2001).

More recently in order to investigate the functional form of the universal
multiplicity function, Yahagi, Nagashima \& Yoshii (2004) (hereafter YNY) performed five runs of N-body simulations
with high mass resolution and compared them with different multiplicity function and with a fit by them proposed.

They showed that discrepancies are observed
%even if there is a good agreement
between some of the quoted analytical multiplicity function with simulations.
%some discrepancies are observed:
%Thet showed that
%
%for example the maximum value of the multiplicity function from their simulations at $\nu \simeq 1$, where $\nu=\left(\frac{\delta_{\rm c}(t)}{\sigma(M)}\right)^2$, is 
%smaller, and its low
%mass tail is shallower when compared with the ST, ST1 multiplicity function.
%

In the present paper, I shall use an improved version of the barrier shape obtained in Del Popolo \& Gambera (1998), obtained from the parameterization of the nonlinear collapse discussed in that paper, taking account of
asphericity and tidal interaction between protohaloes and the effects of a non-zero cosmological constant, together with the results of ST, ST1 in order to study the ``unconditional" multiplicity function.

The reasons that motivates this study are several:\\

%%%%%a)
As previously reported, multiplicity functions like ST and J01, fit only approximatively high resolution N-body
simulations like those of YNY, while
the functional form proposed in YNY, provides a better fit when compared
with the ST functional form. Unfortunately the functional form for the multiplicity function proposed in YNY
%Similarly to the functional form proposed by YNY,
and similarly that of J01 (which is a fit to their ``Hubble Volume" simulations of $\tau$CDM and $\Lambda$CDM cosmologies)
are not based on any theoretical background. So it is important to find a better analytical form, which  starting by ``first principles" is able to fit in a better way simulations and is physically motivated.
I show that the function obtained in the present paper, similarly to that in YNY provides a better fit than the ST or other functional forms used in literature and moreover
%at the same time
it has been obtained from solid physical,
theoretical, arguments.

The paper is organized as follows: in Sect. ~2, I calculate the ``unconditional" multiplicity function.
%and I compare them with previous results and numerical simulations.
%In Sect. ~3, I do the same for the ``conditional" mass function.
% comparing it with numerical simulations.
Sect. ~3 and 4 are devoted to results and to conclusions, respectively.

\section{The barrier model and the multiplicity function}

According to hierarchical scenarios of structure formation, a region collapses
at time $t$ if its overdensity at that time exceeds some threshold. The linear
extrapolation of this threshold up to the present time is called a barrier, B. A
likely form of this barrier is (ST, ST1):
\begin{equation}
B(\sigma ^{2},z)=\sqrt {aS_{\ast}}\left [1+\beta\,\left (S/a S_{\ast}\right )^{\alpha}
\right]=\sqrt{a}\delta _{c}(z)\left[ 1+\frac{\beta }{\left( a\nu \right) ^{\alpha }}\right]
\label{eq:barr}
\end{equation}
In the above equation $a$, $\beta$ and $\alpha$ are constants, $S_{\ast}=\delta_{\rm c}^2$, where $\delta_{\rm c}(t)$
is the linear extrapolation up to the present day of the initial overdensity of
a spherically symmetric region, that collapsed at time t. Additionally,
$S\equiv S_{\ast }\left( \frac{\sigma }{\sigma _{\ast }}\right)^{2}=\frac{S_{\ast }}{\nu}$, 
%$\sigma _{\ast }=\sqrt{S_{\ast }}=\delta _{co}$, 
$\sigma _{\ast }=\sqrt{S_{\ast }}$, 
$\nu=\left(\frac{\delta_{\rm c}(t)}{\sigma(M)}\right)^2$
%$S=\sigma^2(M)$,
where $\sigma^2(M)$ is the present day mass dispersion on comoving scale
containing mass $M$. $S$ depends on the assumed power spectrum. The spherical
collapse model (SC) has a barrier that does not depend on the mass (eg.
Lacey \& Cole 1993 (LC93)). For this model the values of the parameters are $a = 1$ and $\beta= 0$.
The ellipsoidal collapse model (EC)
of ST has a barrier that depends on the mass (moving barrier). The values
of the parameters are $a = 0.707$, $\beta = 0.485$, $\gamma = 0.615$ and are adopted
either from the dynamics of ellipsoidal collapse or from fits to the results of
N-body simulations.

In the following, I shall use an improved version of the barrier obtained in Del Popolo \& Gambera (1998) to get the mass functions, which shall
be compared with those obtained by PS, ST, J01, YNY,
and with numerical simulations of YNY. Since the way the barrier is obtained is described in previous papers (see Del Popolo \& Gambera 1998, 1999, 2000) the reader is referred to those papers for details.
%I'll summarize the way the barrier described in Del Popolo \& Gambera (1998) was obtained.
Assuming that the barrier is proportional to the threshold for the collapse,
similarly to ST, the barrier can be expressed, in the case of a zero cosmological constant, in the form:
\begin{equation}
B(M)=\delta _{\rm c}(\nu,z)=\delta _{\rm co}\left[ 1+
\int_{0}^{r_{\rm ta}}  \frac{r_{\rm ta} l^2 \cdot {\rm d}r}{G M r^3}
\right] \simeq \delta _{\rm co} \left[
1+\frac{\beta_1}{\nu^{\alpha_1}}
\right]
\label{eq:ma7}
\end{equation}
where $\delta _{\rm co}=1.68$ is the critical threshold for a spherical model,
$r_{\rm i}$ is the initial radius, $r_{\rm ta}$ is the turn-around radius,
$l$ the specific angular momentum, $\alpha_1=0.585$ and $\beta_1=0.46$.
%(for $\nu>0.1)$.
The specific angular momentum appearing in Eq. ~(\ref{eq:ma7}) is the specific total angular momentum acquired by the proto-structure during evolution. In order to calculate $L$, I shall use the same model
as described in Del Popolo \& Gambera (1998, 1999) (more hints on
the model and some of the model limits can be found in
Del Popolo, Ercan \& Gambera 2001, Sec. 3).

Assuming a non-zero cosmological constant Eq. (\ref{eq:ma7}) is changed as follows (see Appedix):
\begin{equation}
B(M)=\delta _{\rm c}(\nu,z)=\delta _{\rm co}\left[ 1+
\int_{0}^{r_{\rm ta}}  \frac{r_{\rm ta} l^2 \cdot {\rm d}r}{G M r^3}+\Lambda \frac{r_{\rm ta} r^2}{6 G M}
\right] \simeq \delta _{\rm co} \left[
1+\frac{\beta_1}{\nu^{\alpha_1}}+\frac{\Omega_{\Lambda} \beta_2}{\nu^{\alpha_2}}
\right]
\label{eq:ma8}
\end{equation}
where $\alpha_2=0.4$ and $\beta_2=0.02$ and $\Omega_{\Lambda}$ is the contribution to the density parameter coming from the cosmological constant.
The values of $\alpha_1$, $\alpha_2$, $\beta_1$ and $\beta_2$ are calculated so that the fit function at extreme right hand side of Eq. (\ref{eq:ma8}) reproduces 
the barrier shape (central part of Eq. (\ref{eq:ma8}) depending on $l$ and $\Lambda$).
ST1 connected the form of the barrier with the form of the
multiplicity function.
%They showed that given a mass element, that is a part of
%an halo of mass $M_0$ at time $t_0$, the probability that at earlier time t, this mass
%element was a part of a smaller halo with mass M, is given by the equation:
As shown by ST1, for a given barrier shape, $B(S)$,
%where $S\equiv S_{\ast }\left( \frac{\sigma }{\sigma _{\ast }}\right)^{2}=\frac{S_{\ast }}{\nu}$ and $\sigma _{\ast %}=\sqrt{S_{\ast }}=\delta _{co}$,
the first crossing distribution is well approximated by:
\begin{equation}
f(S)dS=|T(S)|\exp (-\frac{B(S)^{2}}{2S})\frac{dS/S}{\sqrt{2\pi S}}
\label{eq:distrib}
\end{equation}
where $T(S)$ is the sum of the first few terms in the Taylor expansion of $B(S)$:
\begin{equation}
T(S)=\sum_{n=0}^{5}\frac{(-S)^{n}}{n!}\frac{\partial ^{n}B(S)}{\partial S^{n}}
\label{eq:expans}
\end{equation}
The quantity $Sf(S,t)$ is a function of the variable $\nu$ alone.
%, where $\nu\equiv (\delta_c(t)/\sigma(M))^2$.
Since $\delta_c$ and $\sigma$ evolve
with time in the same way, the quantity $Sf(S,t)$ is independent on time. Setting $2Sf(S,t)=\nu f(\nu)$, one obtains the so-called
multiplicity function $f(\nu)$. The multiplicity function is the distribution of first crossings of  a barrier $B(\nu)$ by independent uncorrelated Brownian
random walks (Bond et al. 1991). That's why the shape of the barrier influences the form of the multiplicity function.

In the excursion set approach, the average comoving number density of haloes of mass $M$
%the often called
the universal or ``unconditional" mass function, $n(M,z)$, is given by:
\begin{equation}
n(M,z)=\frac{\overline{\rho}}{M^{2}}\frac{d\log{\nu }}{d\log M}\nu f(\nu )
\label{eq:universal}
\end{equation}
(Bond et al. 1991), where $\overline{\rho}$ is the background density,
%$\nu=\left(\frac{\delta_{\rm c}(z)}{\sigma(m)}\right)^2$ is
%the ratio between the critical overdensity required for collapse in the spherical model, $\delta_{\rm c}(z)$, to the r.m.s. %density fluctuation $\sigma(m)$, on the scale $r$ of the initial size of the object $m$.
In the case of the ellipsoidal barrier shape given in ST, namely Eq. \ref{eq:barr} of the present paper,
%\begin{equation}
%B(\sigma ^{2},z)=\sqrt{a}\delta _{c}(z)\left[ 1+\frac{\beta }{\left( a\nu \right) ^{\alpha }}\right]
%\end{equation}
the Eqs. ~(\ref{eq:distrib}),(\ref{eq:expans}), give, after truncating the expansion at $n=5$ (see ST):
\begin{equation}
\nu f(\nu)=\sqrt{a \nu / 2 \pi}[1+\beta(a
{\nu}^2)^{-\alpha}g(\alpha)]\exp\left(-0.5a\nu^2[1+\beta(a\nu^2)^{-\alpha}]^2\right)
\label{eq:sstt}
\end{equation}
%\begin{eqnarray}
%f(\nu)&=&\sqrt{a \nu / 2 \pi}[1+\beta(a
%{\nu}^2)^{-\alpha}g(\alpha)]\exp\left(-0.5a\nu^2[1+\beta(a\nu^2)^{-\alpha}]^2\right)
%\nonumber \\
%& &
%\simeq A \left( 1+\frac{0.094}{\left( a\nu \right) ^{0.6}}\right) \sqrt{\frac{a\nu }{2\pi }}
%\exp{\{-a\nu \left[ 1+\frac{0.5}{\left( a\nu \right) ^{0.6}}\right] ^{2}/2\}}
%\label{eq:sstt}
%\end{eqnarray}

where
\begin{equation}
g(\alpha)=
\mid 1-\alpha +\frac{\alpha (\alpha
-1)}{2!}-...-\frac{\alpha(\alpha-1)\cdot \cdot \cdot
(\alpha-4)}{5!} \mid
\end{equation}

If the barrier takes account of the cosmological constant, like in Eq. (\ref{eq:ma8}), using the same method
that lead to Eq. (\ref{eq:sstt}), we have that:
\begin{equation}
%f(\nu )d\nu \simeq 1.1\left( 1+\frac{0.073}{\left( a\nu \right) ^{0.585}}\right) \sqrt{\frac{a\nu }{2\pi }}\exp{\{-a\nu
%\left[ 1+\frac{0.52}{\left( a\nu \right) ^{0.585}}\right] ^{2}/2\}}
%f(\nu )d\nu \simeq A _1 \left( 1+\frac{0.1218}{\left( a\nu \right) ^{0.585}}\right) \sqrt{\frac{a\nu }{2\pi }}\exp{\{-
%0.4019 a\nu \left[ 1+\frac{0.5526}{\left( a\nu \right) ^{0.585}}\right] ^{2}\}}
\nu f(\nu )=A _1 \left( 1+\frac{\beta_1 g(\alpha_1)}{\left( a\nu \right) ^{\alpha_1}}
+\frac{\beta_2 g(\alpha_2)}{\left( a\nu \right) ^{\alpha_2}}
\right) \sqrt{\frac{a\nu }{2\pi }}\exp{\{-a \nu \left[ 1+\frac{\beta_1}{\left( a\nu \right) ^{\alpha_1}}
+\frac{\beta_2}{\left( a\nu \right) ^{\alpha_2}}
\right] ^{2}/2\}}
\label{eq:mia}
%f(\nu )d\nu \simeq 1.21\left( 1+\frac{0.06}{\left( a\nu \right) ^{0.585}}\right) \sqrt{\frac{a\nu }{2\pi }}\exp{\{-a\nu
%\left[ 1+\frac{0.57}{\left( a\nu \right) ^{0.585}}\right] ^{2}/2\}}
%\label{eq:mia}
\end{equation}
Using the values for $\beta$ and $\alpha$ of ST ($a=0.707$, $\delta_{\rm c}(z)=1.686 (1+z)$, $\beta \simeq 0.485$ and $\alpha \simeq 0.615$) in Eq. (\ref{eq:sstt}) we get (ST1):

\begin{equation}
\nu f(\nu)\simeq A_2 \left( 1+\frac{0.094}{\left( a\nu \right) ^{0.6}}\right) \sqrt{\frac{a\nu }{2\pi }}\exp{\{-a\nu \left[ 1+\frac{0.5}{\left( a\nu \right) ^{0.6}}\right] ^{2}/2\}}
\label{eq:sstt1}
\end{equation}
%\begin{equation}
%f(\nu )d\nu=A \left( 1+\frac{0.094}{\left( a\nu \right) ^{0.6}}\right) \sqrt{\frac{a\nu }{2\pi }}\exp{\{-a\nu \left[ 1+\frac{0.5}{\left( a\nu \right) ^{0.6}}\right] ^{2}/2\}}
%\label{eq:sstt}
%\end{equation}
with $A_2 \simeq 1$.
This last result is in good agreement with the fit of the simulated first crossing distribution (ST):
\begin{equation}
\nu f(\nu )d\nu =A_3\left( 1+\frac{1}{\left( a\nu \right) ^{p}}\right) \sqrt{\frac{a\nu }{2\pi }}\exp (-a\nu /2)
\label{eq:ssttt}
\end{equation}
where $p=0.3$, and $a=0.707$.
%or
%\begin{equation}
%\nu f(\nu)=A_4(1+\nu'^{-2p})\sqrt{2/\pi}\nu'exp(-\nu'^2/2)
%\end{equation}
%where $\nu' =\nu\sqrt{a}$.
%
%and the values of the constants
%are: $A=0.322$, $p=0.3$ and $a=0.707$.

The normalization factor $A_3$ has to satisfy the constraint:
\begin{equation}
\int_0^{\infty} f(\nu) d \nu=1
\end{equation}
and as a consequence it is not an independent parameter, but is expressed in the form:
\begin{equation}
A=\left[1+2^{-p} \pi^{-1/2} \Gamma(1/2-p)\right]^{-1}=0.3222
\footnote{
%Note that ST used Eq. \ref{eq:sstt} to compare model and data.
Note, that Eq. \ref{eq:ssttt} gives a better fit to Eq. \ref{eq:sstt} if $A \simeq 0.3$ and $a \simeq 0.79$. Vice versa a smaller value of $a$ ($a \simeq 0.63$) and $A=1.08$ in Eq. \ref{eq:sstt} gives a better fit to Eq. \ref{eq:ssttt} (with $A_1=0.3222$ and $a=0.707$), which was the one ST used to compare model and data.}.
\end{equation}
%ST gave the best-fit parameter values as $A=0.322$, $p=0.3$ and $a=0.707$

%%%%%LE EQUAZIONI SOPRA, 6,7 devono avere le p uguali ecc, e sono diverse. CORREGGERE.

%Thus, given Eqs. ~(\ref{eq:distrib})-(\ref{eq:expans}), it is possible to obtain
%mass function, if the barrier shape and the power spectrum are given.

%Putting Eq. (\ref{eq:ma7}) into Eqs. ~(\ref{eq:distrib})-(\ref{eq:expans}) and truncating the %expansion at $n=5$,

In the case of the barrier given in Eq. (\ref{eq:ma7}), the ``unconditional" multiplicity function can be approximated by:
\begin{equation}
%f(\nu )d\nu \simeq 1.1\left( 1+\frac{0.073}{\left( a\nu \right) ^{0.585}}\right) \sqrt{\frac{a\nu }{2\pi }}\exp{\{-a\nu
%\left[ 1+\frac{0.52}{\left( a\nu \right) ^{0.585}}\right] ^{2}/2\}}
%f(\nu )d\nu \simeq A _1 \left( 1+\frac{0.1218}{\left( a\nu \right) ^{0.585}}\right) \sqrt{\frac{a\nu }{2\pi }}\exp{\{-
%0.4019 a\nu \left[ 1+\frac{0.5526}{\left( a\nu \right) ^{0.585}}\right] ^{2}\}}
\nu f(\nu ) \simeq A _4 \left( 1+\frac{b}{\left( a\nu \right) ^{0.585}}\right) \sqrt{\frac{a\nu }{2\pi }}\exp{\{-a c\nu \left[ 1+\frac{d}{\left( a\nu \right) ^{0.585}}\right] ^{2}\}}
\label{eq:miaa}
%f(\nu )d\nu \simeq 1.21\left( 1+\frac{0.06}{\left( a\nu \right) ^{0.585}}\right) \sqrt{\frac{a\nu }{2\pi }}\exp{\{-a\nu
%\left[ 1+\frac{0.57}{\left( a\nu \right) ^{0.585}}\right] ^{2}/2\}}
%\label{eq:mia}
\end{equation}
where $a=0.707$, $b=0.1218$, $c=0.4019$, $d=0.5526$ and $A_4 \simeq 1.75$ is obtained from the normalization condition.
%IMPORTANTE TOLTO
%%\begin{equation}
%%\nu f(\nu )=\frac{A}{2}\left( 1+\frac{1}{\nu _{1}^{\alpha_2 }}+\frac{1}{\nu _{2}^{\beta_2 }}\right) \sqrt{\frac{\nu %%_{1}}{2\pi }}\exp (-\nu _{3}/2)
%%\end{equation}
%where $A \simeq 0.395$, $\nu _{1}=0.83 \nu$, $\nu _{2}=0.12 \nu$, $\nu _{3}=0.73 \nu$, $\alpha_2=0.3$ and $\beta_2=0.12$.
%%where $A \simeq 0.395$, $\nu _{1}=0.83 \nu$, $\nu _{2}=0.12 \nu$, $\nu _{3}=0.8 \nu$, $\alpha_2=0.355$ and %%$\beta_2=0.028$.

In the case of the barrier with non-zero cosmological constant, Eq. (\ref{eq:ma8}), a good approximation to the multiplicity function is given by:
\begin{equation}
\nu f(\nu ) \simeq A _5 \left( 1+\frac{0.1218}{\left( a\nu \right) ^{0.585}}
+\frac{0.0079}{\left( a\nu \right) ^{0.4}}
\right) \sqrt{\frac{a\nu }{2\pi }}\exp{\{-0.4019 a \nu \left[ 1+\frac{0.5526}{\left( a\nu \right) ^{0.585}}
+\frac{0.02}{\left( a\nu \right) ^{0.4}}
\right] ^{2}\}}
%\label{eq:mia}
\label{eq:mia1}
\end{equation}
where $A_5=1.75$.
As previously reported, for matter of completeness, to the previous functions, namely PS, ST, Eq. (\ref{eq:mia1})
we have to add J01, which satisfies the equation:
\begin{equation}
\nu f(\nu)=0.315 exp(-\mid 0.61+ln[\sigma^{-1}(M)]\mid^{3.8})
\end{equation}
In order to express the above relation as a function of $\nu$, I
substitute $\sigma^{-1}(M)=\nu/\delta_c$ and I assume a
constant value of $\delta_c$, that of the Einstein-de Sitter Universe namely
$\delta_c=1.686$. The above formula is valid for $0.5 \leq \nu \leq  4.8$.\

YNY (Eq. 7, hereafter YNY7) proposed the following function to fit the numerical multiplicity function:
\begin{eqnarray}
%\nu f(\nu) = A [1+(B \nu / \sqrt{2})^C] \nu^D \exp[-(B \nu)^2/2]
\nu f(\nu) = A [1+(B \nu / \sqrt{2})^C] \nu^D \exp[-(B \nu/\sqrt{2})^2],
\label{eq:4fit}
\end{eqnarray}
where, $A$ is a normalization factor to satisfy the unity constraint,
$\int_0^{\infty}f(\nu)d\nu=1$, therefore
\begin{eqnarray}
A=2 (B/\sqrt{2})^D\{\Gamma[D/2] + \Gamma[(C+D)/2]\}^{-1}.
\end{eqnarray}
The best-fit parameters are given
as $B$=0.893, $C$=1.39, and $D$=0.408, and from these parameters, $A$
is constrained so that $A=0.298$.

The CDM spectrum used in the present paper is that of Bardeen et al. (1986)(equation~(G3)).
%, with transfer function:
%\begin{equation}
%T(k) = \frac{[\ln \left( 1+2.34 q\right)]}{2.34 q}
%\cdot [1+3.89q+
%(16.1 q)^2+(5.46 q)^3+(6.71)^4]^{-1/4}
%\label{eq:ma5}
%\end{equation}
%where
%$q=\frac{k\theta^{1/2}}{\Omega_{\rm X} h^2 {\rm Mpc^{-1}}}$.
%Here $\theta=\rho_{\rm er}/(1.68 \rho_{\rm \gamma})$
%represents the ratio of the energy density in relativistic particles to
%that in photons ($\theta=1$ corresponds to photons and three flavors of
%relativistic neutrinos).
%The power spectrum was normalized to reproduce the observed abundance of rich
%cluster of galaxies (e.g., Bahcal \& Fan 1998).
%

\section{Results}

In this section, I compare the analytic multiplicity functions of PS, ST, J01, YNY7, and Eq. (\ref{eq:mia1}), of the present paper, with the numerical simulations
of YNY.
%We used the Adaptive Mesh Refinement $N$-body code developed by
%\citet{yahagi}, which is a  vectorized and parallelized version of the
%code described in \citet{yy01}.
%All five runs of
Those simulations adopt the $\Lambda$CDM cosmological parameters of $\Omega_m=0.3$,
$\Omega_\lambda=0.7$, $h=0.7$, and $\sigma_8=1.0$, using
$512^3$ particles in common (see YNY for details).

The comparison between numerical multiplicity functions and theoretical ones
is shown in Fig. 1.

In the plot the solid line represents the multiplicity function
obtained in the present paper, the short-dashed
line YNY7, the dotted line the ST multiplicity function,
the long-dashed line the J01 multiplicity function. The errorbars with open
circles represents the run 140 of YNY, those with filled squares the case 70b,
those with open squares the case 70a, those with filled circles the case 35b,
those with crosses the case 35a.

%This situation is more clearly visible in Fig. 2, where J01 ({long-dashed line}), ST ({dotted line}), YNY7 ({short-dashed %line}) and Eq. (\ref{eq:mia1}) ({solid line}) of the present paper, are compared.

Since the data are available only in the region at
$\nu \leq 3$, these functions could be erroneous at $\nu \geq 3$.

Note that the comparison of the above curves, except  for the PS model, with the results of N-body simulations show a very good agreement.
%Such a comparison is
%presented by  Yahagi et al. (2004).
%According to their results,
However, there are some discrepancies between the YNY
multiplicity function and other model functions (except this in the present paper).
%First, in the low $\nu$
%However, some detailed discrepancies are seen between
First, the multiplicity function of the present paper, similarly to that of YNY,
%and the ST and J01 analytic
%functions.
%First,
in the low-$\nu$ region of $\nu \leq 1$,
%our
%numerical multiplicity functions
systematically falls below the ST and
the J01 functions. In this region the multiplicity function of the present paper is very close to that of YNY.

As seen in Fig. 1, and in agreement with YNY, the numerical multiplicity functions reside between the ST and J01 multiplicity functions at
%$\nu \geq 3$
$2 \leq \nu \leq 3$ (except for the run 35b).
%and are below the ST function at $\nu \leq 1$.
Additionally, the numerical multiplicity functions have an apparent peak at $\nu \sim 1$  instead of the plateau
that is seen in the J01 function.

%, while they are consistent with that of
%\citetalias{wht02}.
On the other hand, in the high-$\nu$ region, where
%%%%%%%%%%%%%%%%%%%%%%%%%%%%%%%%%%%%%%%%%%%%%%%%%%
$\nu$ is significantly larger
than unity, the multiplicity function of the present paper
%take values between the YNY and J01 functions
%(POI E' SOPRA..)
%while
like YNY takes values between ST
and J01 functions. These differences between numerical multiplicity functions and analytic ones, like ST, ST1 and J01,
are within 1 $\sigma$ error bars, and they are possibly due
to the different box sizes adopted (see YNY for a discussion). To be more precise, throughout the peak range of $0.3 \leq \nu \leq 3$, the ST multiplicity function is in disagreement with the high mass resolution $N$-body simulations of YNY and that of the present paper. As shown by YNY the ST functional form provides a good fit to them only choosing  parameter values of $a=0.664, p=0.321$, and
$A_3=0.301$. The multiplicity function obtained in the present paper has a peak at $\nu \sim 1$ as in the ST function,
and YNY numerical multiplicity function and YNY7, instead of a plateau as in the J01 function.

%\citet{st99} used the data from the
%GIF simulations \citep{gif99} with the box size of 144$h^{-1}$Mpc or less,
%and \citetalias{wht02} mainly used the data from the box size of
%200$h^{-1}$Mpc.  On the other hand, \citetalias{j01} used the
%3000 $h^{-1}$Mpc simulation at maximum.  We have taken into account
%the box-size effect using the conditional PS multiplicity function
%(equation \ref{eq:cndPS})
%but this effect is found to be too small to resolve this problem.
%Introducing the estimation using the conditional multiplicity
%function based on the unconditional multiplicity function which fits
%the numerical multiplicity function well, such as the ST multiplicity
%function, would resolve this problem. However, such an improved
%estimation of the box-size effect might be weaker than
%the estimation based on the PS function, because
%the unconditional ST function at $\nu\sim1$ has a broad
%peak that is below that of the PS function.  The fact that our
%numerical multiplicity functions keep the universality supports this
%line of argument.

%Thus, there are two discrepancies remained.  First one is the
%discrepancy between the numerical multiplicity function and the model
%numerical multiplicity function.  We require a new model function
%which fit numerical multiplicity function better keeping some required
%conditions such as, $\int_0^{\infty}f(\nu)\mbox{d}\nu=1$.

%%%%A second discrepancy is observed between the numerical and analytical multiplicity functions.
%Thus, there are two discrepancies remained.  One is the
%discrepancy between the numerical and analytical multiplicity functions.
I want to stress that the functional form proposed in YNY, namely YNY7, provides a better fit when compared
with the ST functional form but
%we need an analytic function based on a
it is not based on theoretical background. The function obtained in the present paper, similarly to YNY7 provides a better fit
to simulations
than the ST functional form, and at the same time has been obtained from solid physical, theoretical, arguments.
%which fits the numerical multiplicity function
%even better.
The better agreement observed between the multiplicity function of the present paper and YNY simulations, when compared with the ST,
is connected to the shape of the barrier ($\delta_{\rm c}$).
%As reported in Sec. 2,
%In the spherical collapse model, this critical density does not depend on the mass of the collapsed object.
Taking account of the effects of asphericity and tidal interaction with neighbors, Del Popolo \& Gambera (1998),
%using a parametrization of the ellipsoidal collapse,
showed that the threshold is mass dependent, and in particular that of the set of
objects that collapse at the same time, the less massive ones must initially have been denser than the more massive,
since the less massive ones would have had to hold themselves together against stronger tidal forces.

The shape of the barrier given in Eq. (\ref{eq:ma7}) is a direct consequence of the angular momentum acquired by the proto-structure during evolution while Eq. (\ref{eq:ma8}) introduces the effects of the cosmological constant.
%The good agreement between Sheth \& Tormen (1999) model and that of the present paper is due to the similitude of the
%barriers of the two papers.
%In both two,

Similarly to ST, the barrier increases with $S$ (decrease with mass, M) differently from other models (see Monaco 1997a, b).
It is interesting to note that the increase of the barrier with $S$ has
several important consequences and these models have a richer structure
than the constant barrier model.
%In the case of non-spherical collapse
%with
%increasing
%barrier, a small fraction of
%the mass in the universe remains unbound, while for the spherical dynamics, at the given time, all the mass is bound
%up in collapsed objects. Moreover, incorporating the non-spherical collapse
%with increasing
%barrier in the excursion set approach results in a model in which
%fragmentation and mergers may occur (ST). If the barrier decreases with
%$S$ (Monaco 1997 a,b), this implies that all walks are guaranteed
%to cross it and so there is no fragmentation associated with this
%barrier shape.

\begin{figure}
\centerline{\hbox{
\psfig{file=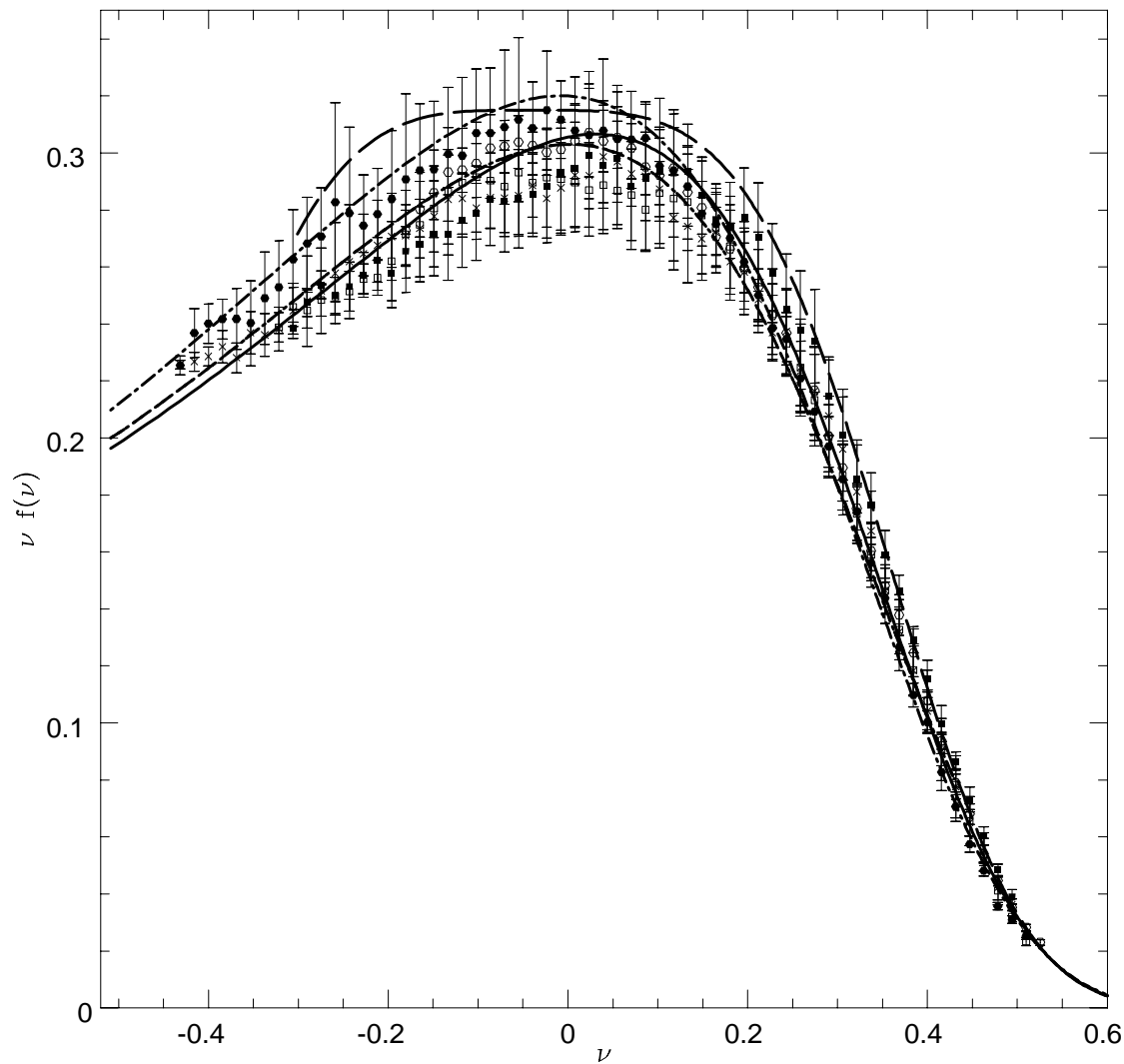,width=16cm}
}}
\caption[]{The best-fit multiplicity function.
%equation \ref{eq:4fit}
%({\it solid line}).
In the plot the solid line represents the multiplicity function
obtained in the present paper, the short-dashed
line YNY7, the dotted line the ST multiplicity function,
the long-dashed line the J01 multiplicity function. The errorbars with open
circles represents the run 140 of YNY, those with filled squares the case 70b,
those with open squares the case 70a, those with filled circles the case 35b,
those with crosses the case 35a
}
\end{figure}

\begin{figure}
\centerline{\hbox{
\psfig{file=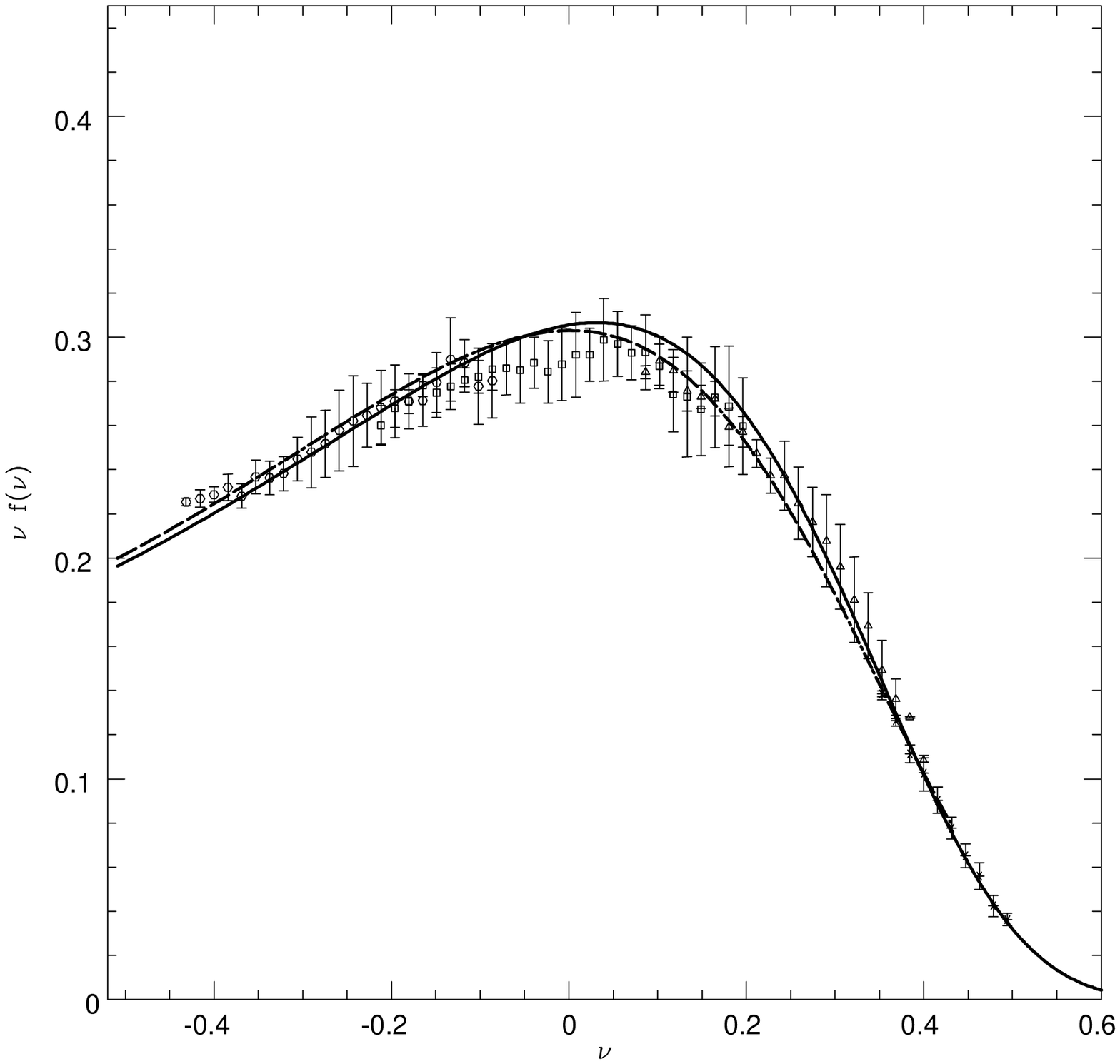,width=16cm}
}}
\caption[]{Time dependence of the multiplicity function
%. Theshows the multiplicity
%function
from the 35a run, for four redshift ranges of
$0 \leq z < 1$ ({open circles}), $1 \leq z < 3$ ({open squares}),
$3 \leq z < 6$ ({open triangles}), and $z \geq 6$,    ({crosses}).
Also shown are
%the best fit function for all the runs using the ST
%functional form
%({\it dot-dashed line})
YNY7 (solid line)
%and  equation \ref{eq:4fit} ({\it solid line}).
and the model of the present paper (dot-dashed line).
}
\end{figure}

%In other words, this

%{\bf The barrier given in Eqs. ~(\ref{eq:ma7})-(\ref{eq:ma8}), differently from that of the spherical collapse is mass %dependent.
%Eqs. ~(\ref{eq:ma7})-(\ref{eq:ma8}) show
%that the threshold
%for collapse decreases with mass, or similarly it increases with $\sigma$ since this quantity is a decreasing function of %mass.
%%
%%In other words, this means that, in order to form
%%structure, peaks in more dense
%%regions must have a lower value of the threshold, $\delta_c(nu)$, with respect
%%to those of under-dense regions.}
%}
The decrease of the barrier with mass means that, in order to form structure, more massive peaks must
cross a lower threshold, $\delta_c(\nu,z)$, with respect to under-dense ones.
At the same time, since the
probability to find high peaks is larger in more dense regions,
this means that, statistically, in order to form structure,
peaks in more dense
regions may have a lower value of the threshold, $\delta_c(\nu,z)$, with respect
to those of under-dense regions.
This is due to
the fact that less massive objects are more influenced by external tides, and
consequently they must be more overdense to collapse by a given time.
In fact, the angular momentum acquired by a shell centred on a peak
in the CDM density distribution is anti-correlated with density: high-density
peaks acquire less angular momentum than low-density peaks
(Hoffman 1986; Ryden 1988).
A larger amount of angular momentum acquired by low-density peaks
(with respect to the high-density ones)
implies that these peaks can more easily resist gravitational collapse and consequently
it is more difficult for them to form structure.
%
%%This is in agreement with Audit et al. (1997), Peebles (1990) and Del Popolo \& Gambera (1998),
%%which pointed out that the gravitational collapse is slowed down by the  effect of the shear
%%rather than fastened by it (as substained by other authors).
%
Therefore, on small scales, where the shear is statistically greater,
structures need, on average, a higher  density contrast to collapse.

It is evident that the effect of a non-zero cosmological
constant adds to that
%is that of reducing the effect
of L.
%\footnote{I studied in a previous paper the effect of a non-zero cosmological constant on evolution
%of some relations like the M-T relation (Del Popolo 2002).
%The evolution of the M-T relation is more rapid in models
%with L =0.
The effect of a non-zero cosmological constant is that of
slightly changing the evolution of the multiplicity function with respect to
open models with the same value of $\Omega_0$. This is caused by the fact that
in a flat universe with $\Omega_{\Lambda}>0$, the density of the universe remains close
to the critical value later in time, promoting perturbation growth
at lower redshift. The evolution is more rapid for larger values (in
absolute value) of the spectral index, n.
%}.
%and that the CDM
%cosmology, with 0 =0.3,  =0.7, is in better agreement with
%the observed bend than the 0 =0.3 OCDM model.

As previously reported, the ST model gives a better fit to simulations than PS model, but
it has some discrepancies with simulations.
ST model was introduced at the beginning (Sheth \& Tormen 1999) as a fit to the GIF simulations and in  a subsequent paper (SMT) was recognized the importance of aspherical collapse in the
functional form of the mass function. The effects of asphericity were taken into account by changing the functional form of the critical overdensity (barrier) by means of a simple intuitive parameterization of elliptical collapse of isolated spheroids. The model proposed in the present paper has several similitudes with ST and ST1 models, namely it uses the excursion
set approach as extended by ST1 to calculate the multiplicity function, but at the same time it differs from ST and ST1 for the way the barrier was calculated and for the fact that takes account of angular momentum acquisition, and a non-zero cosmological constant,
things which are not
%that is not
taken into account into ST and ST1. These differences gives rise to a multiplicity function in better agreement with simulations.
This shows the importance of the form of the barrier.
%: in the case of the PS model it was constant, describing a spherical collapse. In ST, it was
%mass dependent and taking account the effects of ellipsoidal collapse, with a noteworthy
%improvement in the multiplicity function.
%The quoted discrepancy (PS model versus N-body simulations and ST model)
%is not surprising since the PS model, as any other analytical model, should make several %assumptions to get simple
%analytical predictions. As previously reported, the main assumptions that the PS model combines %are the simple physics
%of the spherical collapse model with the assumption that the initial fluctuations were Gaussian %and small.
%
%The above considerations show that it is possible to get accurate predictions for a number of statistical quantities
%associated with the formation and clustering of dark matter haloes by incorporating a non-spherical collapse in the
%excursion set approach.
The improvement of the model of the present paper and ST model with respect to PS is probably connected also to the fact that incorporating the non-spherical collapse with increasing barrier in the excursion set approach results in a model in which fragmentation and mergers may occur, effects important in structure formation. In the case of
non-spherical collapse with increasing barrier, a small fraction of
the mass in the Universe remains unbound, while for the spherical
dynamics, at the given time, all the mass is bound up in collapsed
objects. Moreover, incorporating the non-spherical collapse with
increasing barrier in the excursion set approach results in a model
in which fragmentation and mergers may occur (ST). If the barrier
decreases with S (Monaco 1997a,b), this implies that all walks are
guaranteed to cross it and so there is no fragmentation associated
with this barrier shape.

In other words, the excursion set approach with a barrier taking account effects of physics of
structure formation gives rise to good approximations to the numerical multiplicity function: the  approximation goodness increases with a more improved form of the barrier (taking account more and more physical effects: angular momentum acquisition, non zero cosmological constant, etc).
Another important aspect of the quoted method is its noteworthy versatility: for example it is very easy to take account of the presence of a non zero
cosmological constant englobing it in the barrier. I recall that the YNY numerical multiplicity function assumes a non zero cosmological constant while the theoretical models (ST,ST1, J01) does not take this into account.
%Following the same method of Del Popolo \& Gambera (1998) it is easy to englobe the
%cosmological constant effects. We shall show this in a future paper.

%In most cases, the numerical multiplicity functions and the best-fit
%functions to them are consistent with the ST and J01 multiplicity
%functions at $\nu \gtrsim 3$.  However, each of the numerical multiplicity
%functions reside between the ST and J01 functions at $1.5
%\lesssim \nu \lesssim 3$, and is below the ST
%function at $\nu \lesssim 1$ except for the 35b run.  The
%numerical multiplicity functions have an apparent peak at $\nu
%\sim 1$, instead of a plateau as seen in the J01 function.

Finally I checked the time dependence of the multiplicity function.
Fig. 2 shows the multiplicity
function from the 35a run, for four redshift ranges of
$0 \leq z < 1$ ({open circles}),
$1 \leq z < 3$ ({open squares}),
$3 \leq z < 6$ ({open triangles}), and
$z \geq 6$,    ({crosses}).
%The dashed line represents YNY7 while the solid line Eq. (\ref{eq:mia}) of te present paper.
At high redshifts, high-$\nu$ halos in the exponential part of the
%best-fit ST function and equation \ref{eq:4fit}
YNY7 (solid line) function and Eq. (\ref{eq:mia1}) (dot-dashed line of the present paper)
are probed.  As redshift decreases,
the probe window moves to the lower-$\nu$ region.
Fig. 2 shows that the multiplicity function of this paper, Eq. (\ref{eq:mia1}), and YNY7 both gives a good fit to the numerical simulations. For small values of $\nu$, Eq. (\ref{eq:mia1}) is a slightly better fit to data, and at large values of $\nu$ the two functions decays in the same way.

\section{Conclusions}

In the present paper, I compared the numerical multiplicity function given in YNY with the theoretical multiplicity function obtained by means of the excursion set model and an improved version of the barrier shape obtained in Del Popolo \& Gambera (1998), which implicitly takes account of tidal interactions between clusters and a non-zero cosmological constant.
I showed that the barrier obtained in Del Popolo \& Gambera (1998)
%, which takes account of asphericity and tidal interaction
%between proto-haloes,
gives rise to a better description of the multiplicity functions than other models (ST, J01) and the
agreement is based on sound theoretical models and not on fitting to simulations.

The main results of the paper can be summarized as follows: \\
1) the non-constant barrier of the present paper
%obtained from the non-spherical collapse in Del Popolo \& Gambera (1998), taking account of the %tidal interaction of proto-clusters with neighboring ones,
combined with the ST1 model gives ``unconditional" multiplicity functions in better agreement with the N-body simulations of YNY
than other previous models (ST, ST1, J01). \\
2) The comparison of the theoretical multiplicity function of the present paper, in agreement with the YNY result,
shows some discrepancies with the theoretical multiplicity function of several authors (ST, ST1, J01): e.g., the maximum value of the
multiplicity function from  simulations at $\nu \sim 1$ is smaller, and its low mass tail is shallower when
compared with the ST multiplicity function.\\
%, in agreement with the results of YNY.
3) The multiplicity function of the present paper gives a good fit to simulations results as the fit function proposed by YNY, but differently from that it was obtained from a
sound theoretical background.\\
4) The excursion set model with a moving barrier is very versatile since it is very easy to introduce easily several physical effects in the
calculation of the multiplicity function, just modifying the barrier.\\
%%%%5) The ST formulae really do work  for different barrier shapes (at least that used in this paper, that introduced in SMT and that of Monaco (1997a,b)).\\
%%%%6) The behavior of the ``unconditional" mass function at small masses is similar to that of ST, ST1, and very different from that proposed by J01 (see ST Fig. 13).

The above considerations show that it is possible to get accurate predictions for a number of statistical quantities associated with the formation and clustering of dark matter haloes by incorporating a non-spherical collapse which
takes account of a non-zero cosmological constant in the
excursion set approach. The improvement is probably connected also to the fact that incorporating the non-spherical collapse with increasing barrier in the excursion set approach results in a model in which fragmentation and mergers may occur, effects important in structure formation.
Moreover, the effect of a non-zero cosmological
constant adds to that
%is that of reducing the effect
of angular momentum
%\footnote{I studied in a previous paper the effect of a non-zero cosmological constant on evolution
%of some relations like the M-T relation (Del Popolo 2002).
%The evolution of the M-T relation is more rapid in models
%with L =0.
%
%The effect of a non-zero cosmological constant is that of
slightly changing the evolution of the multiplicity function with respect to
open models with the same value of matter density parameter.
%This is caused by the fact that
%in a flat universe with $\Omega_{\Lambda}>0$, the density of the universe remains close
%to the critical value later in time, promoting perturbation growth
%at lower redshift. The evolution is more rapid for larger values (in
%absolute value) of the spectral index, n.

%

\section{Appendix}

The equation governing the collapse of a density perturbation taking account
angular momentum acquisition by protostructures
can be obtained using a model due to Peebles (Peebles 1993) (see also Del Popolo \& Gambera 1998, 1999).\\
 Let's consider an ensemble of gravitationally growing mass concentrations
and suppose that the material in each system collects within the
same potential well
with inward pointing acceleration given by $g(r)$ (see Del Popolo \& Gambera 1998). We
indicate with $dP=f(L,r v_r,t)dL dv_r dr$ the probability that a particle, of mass $m$,
can be found  in the proper radius range $r$, $r+dr$, in the radial
velocity range $v_r={\dot r}$, $v_r+d v_r$ and with angular momentum
$L=m r v_\theta$ in the range $dL$, or specific angular momentum $l=L/m=r v_\theta$.
%then from Liouville's theorem it follows that
%the distribution function, $f$,
%satisfies the collisionless Boltzmann equation:
%\begin{equation}
%\frac{\partial f}{\partial t} + v_{r}
%\frac{\partial f}{\partial r} + \frac{\partial f}{\partial v_{r}}
%\cdot [ \frac{L_{2}}{r^{3}} - g(r)] = 0
%\end{equation}
The radial
acceleration of the particle is:
\begin{equation}
\frac{dv_r}{dt}=\frac{l^2(r)}{r^3}-g(r)=\frac{l^2(r)}{r^3}-\frac{G M }{r^2}
\label{eq:col}
\end{equation}
%%where $g(r)$ is the acceleration.
%and $g(r)$ the acceleration.
where $M$ is the mass of the central concentration.  
Eq. (\ref{eq:col}) can be derived from a potential
and then from Liouville's theorem it follows that
the distribution function, $f$,
satisfies the collisionless Boltzmann equation:
\begin{equation}
\frac{\partial f}{\partial t} + v_{r}
\frac{\partial f}{\partial r} + \frac{\partial f}{\partial v_{r}}
\cdot \left[ \frac{l^{2}}{r^{3}} - g(r) \right] = 0
\end{equation}
%The equation governing the collapse of a density perturbation

Assuming a non-zero cosmological constant Eq. (\ref{eq:col})
becomes:
%taking account of a cosmological constant and
%the
%interaction of the quadrupole moment of the system with the tidal field of
%the matter of the neighbouring proto-clusters is given by:
\begin{equation}
\frac{dv_r}{dt}=-\frac{G M }{r^2}+\frac{l^2(r)}{r^3}+\frac{\Lambda}{3} r \label{eq:coll}
\end{equation}
(Peebles 1993; Bartlett \& Silk 1993; Lahav 1991; Del Popolo \& Gambera 1998, 1999).
%where, $L(r)$ is the angular momentum.
Integrating Eq. (\ref{eq:coll}) we have:
\begin{equation}
\frac{1}{2}\left( \frac{dr}{dt}\right) ^{2}=\frac{GM}{r}+\int
\frac{l^{2}}{r^{3}}dr+\frac{\Lambda }{6}r^{2}+\epsilon
\label{eq:coll1}
\end{equation}
where the value of the specific binding energy of the shell, $\epsilon$, can be obtained using the condition for turn-around, $\frac{dr}{dt}=0$.

In turn the binding energy of a growing mode solution is uniquely given
by the linear overdensity, $\delta_{i}$, at time $t_{i}$.
From this overdensity, using the linear theory, we may obtain that of
the turn-around epoch and then that of the collapse.
%which is given by:
We find the binding energy of the shell, $C$, using the
relation between $v$ and $\delta_{i}$ for the growing mode
(Peebles 1980) in Eq. (\ref{eq:coll1}) and finally the
linear overdensity at the time of collapse is given by:
%Using
%a technique similar to that by Bartlett \& Silk (1993)
%it is possible to obtain the overdensity at the time of collapse:
\begin{equation}
\delta _{\rm c}=\delta _{\rm co}\left[ 1+
\int_{0}^{r_{\rm ta}}  \frac{r_{\rm ta} l^2 \cdot {\rm d}r}{G M r^3}+\Lambda \frac{r_{\rm ta} r^2}{6 G M}
\right] \simeq \delta _{\rm co} \left[
1+\frac{\beta_1}{\nu^{\alpha_1}}+\frac{\Omega_{\Lambda} \beta_2}{\nu^{\alpha_2}}
\right]
\label{eq:maa}
\end{equation}
where $\alpha_1=0.585$, $\beta_1=0.46$, $\alpha_2=0.4$ and $\beta_2=0.02$

\end{document}